\documentclass[aps,prl,showpacs,twocolumn,superscriptaddress]{revtex4}
\usepackage{amsmath}
\usepackage{amssymb}
\usepackage{graphicx}
\begin{document}
\title{Striped Multiferroic Phase in Double-Exchange Model for Quarter-Doped Manganites}
\author{Shuai Dong}
\affiliation{Department of Physics and Astronomy, University of Tennessee, Knoxville, Tennessee 37996, USA}
\affiliation{Materials Science and Technology Division, Oak Ridge National Laboratory, Oak Ridge, Tennessee 32831, USA}
\affiliation{Nanjing National Laboratory of Microstructures, Nanjing University, Nanjing 210093, China}
\author{Rong Yu}
\affiliation{Department of Physics and Astronomy, University of Tennessee, Knoxville, Tennessee 37996, USA}
\affiliation{Materials Science and Technology Division, Oak Ridge National Laboratory, Oak Ridge, Tennessee 32831, USA}
\author{J.-M. Liu}
\affiliation{Nanjing National Laboratory of Microstructures, Nanjing
University, Nanjing 210093, China} \affiliation{International Center
for Materials Physics, Chinese Academy of Sciences, Shenyang 110016,
China}
\author{Elbio Dagotto}
\affiliation{Department of Physics and Astronomy, University of Tennessee, Knoxville, Tennessee 37996, USA}
\affiliation{Materials Science and Technology Division, Oak Ridge National Laboratory, Oak Ridge, Tennessee 32831, USA}
\date{\today}

\begin{abstract}
The phase diagram of quarter-hole-doped perovskite manganites is investigated using the
double-exchange model. An exotic striped type-II multiferroic phase, where $25\%$ of the
nearest-neighbor spin couplings are orthogonal to each other, is found in the
narrow-bandwidth region. Comparing with the spiral-spin ordering phase of undoped
manganites, the multiferroic Curie temperature of the new phase is estimated to be
$\sim4$ times higher, while the ferroelectric polarization
is similar in magnitude. Our study provides a path for
non-collinear spin multiferroics based on electronic self-organization,
different from the traditional approach based on superexchange
frustration.
\end{abstract}
\maketitle

\textit{Introduction}. Multiferroics, materials where ferroelectric (FE) and magnetic
orders coexist, have attracted enormous interest due to
their technological relevance and fundamental science challenges \cite{Fiebig:Jpd}.
Based on the microscopic origin of the FE polarization ($P$), the multiferroics can be
classified into two families \cite{Khomskii:Phy}. Type-I multiferroics, where
ferroelectricity and magnetism have different origins, are often good ferroelectrics
with high FE transition temperatures and large $P$. However, the
coupling between magnetism and ferroelectricity is usually weak. In contrast,
in the type-II multiferroics (magnetic multiferroics), such as $R$MnO$_3$
($R$=Tb and Dy) \cite{Kimura:Nat}, the ferroelectricity is caused by, and thus it is
strongly coupled with, a particular magnetic order. The pursuit for type-II
multiferroics with higher critical temperatures $T_{\rm C}$ and larger $P$
is one of the most fundamental challenges of modern condensed-matter physics~\cite{Kimura:Nm}.

The manganite family provides fertile ground for both type-I (e.g.
BiMnO$_3$ and hexagonal YMnO$_3$) and type-II (e.g. $R$MnO$_3$ perovskites, $R$=Tb, Dy, Ho,
or Eu$_{1-x}$Y$_x$) multiferroics. Recent theoretical studies predicted
additional type-I multiferroics in manganites, e.g. Pr$_{0.6}$Ca$_{0.4}$MnO$_3$, CaMnO$_3$,
and BaMnO$_3$ \cite{Efremov:Nm}. However,
since among type-II multiferroics the manganite multiferroics show the largest
$P$ (e.g. $\sim0.6$-$2.5\times10^3$ $\mu$C/m$^2$ for TbMnO$_3$ and
DyMnO$_3$ \cite{Kimura:Nat}, $\sim10^4$ $\mu$C/m$^2$ for HoMnO$_3$ \cite{Sergienko:Prl}),
it would be even more exciting if new type-II manganite multiferroics with higher $T_{\rm C}$
could be discovered.

Although the manganite phase diagrams
have been extensively studied \cite{Dagotto:Prp},
exotic new phases may still be waiting to be discovered,
particularly in the narrow-bandwidth limit where the
multiferroic manganites are located.
Therefore,
it is important to continue the detailed investigation of this
narrow-bandwidth regime.
Along this direction, recent studies addressed
the systematic phase diagram for undoped
manganites using the two-orbital double-exchange (DE)
model \cite{Dong:Prb08.2}.

Compared with the undoped case, the doped multiferroics, which have not been widely
studied, can provide a fruitful playground since one extra degree of freedom,
the charge, is now active. It is already known that doping-driven electronic self-organization can be very important to the emergence of high-$T_{C}$ superconductivity and colossal magnetoresistance \cite{Dagotto:Sci}. Thus,
it is natural to explore a
similar path in multiferroics. For instance, very recently, the Ca-doped BiFeO$_3$ and TbMnO$_3$ were studied experimentally \cite{Yang:Nm,Mufti:Prb}. In this Letter, theoretical investigations are extended
to the doped manganites, revealing a type-II multiferroic phase at quarter-doping  that was missing in previous theoretical studies~\cite{Hotta:Prl}.

\textit{Model and methods}.
In this letter, the phase diagram of quarter-doped manganites will be studied
using the two-orbital DE model. The Hamiltonian is:
\begin{equation}
H=H_{\rm DE}(t_0)+H_{\rm SE}(J_{\rm AF})+H_{\rm lattice}(\lambda),
\end{equation}
which includes the two-orbital DE hopping term $H_{\rm DE}$,
the nearest-neighbor (NN) superexchange (SE) energy $H_{\rm SE}$, and the lattice
contribution $H_{\rm lattice}$ that contains the electron-phonon coupling
and lattice's elastic energy.
Details of this well-known Hamiltonian can be found
in previous publications \cite{Dagotto:Prp,Dong:Prb08.2}.
The energy unit will be the DE hopping $t_0$ ($\sim0.3$-$0.5$ eV).
Besides $t_0$, there are two main parameters in this Hamiltonian: $J_{\rm AF}$
for the SE, and $\lambda$ for the electron-phonon coupling.

Three numerical methods have been employed to cross check the results. The first
technique is the zero temperature ($T$) variational method where the ground-state energies of several
phases are compared. This calculation can be applied on both finite- and infinite-size lattices.
However, a set of candidate phases must be preselected.
To avoid such a possible bias, the finite-$T$ Monte Carlo (MC) simulation is also here
performed, on finite-size clusters, to confirm the phase diagram obtained by the
variational method. The typical clusters, with periodic boundary conditions, used here
are 2-dimensional (2D) $L\times L$ ($L=8$ and $16$).
Some hidden phases or phase-separation tendencies missed by the first method
can be revealed in the MC simulation. Finally, the zero-$T$ relaxation technique
is used to optimize the ground state spins' and lattice' patterns on a
finite-size cluster. The second and third methods are only applied to 2D lattices
in the current effort because they are computationally intense.
The three processes are cycled several times to improve the accuracy of the
phase diagram. 

\begin{figure}
\vskip -0.9cm
\includegraphics[width=0.50\textwidth]{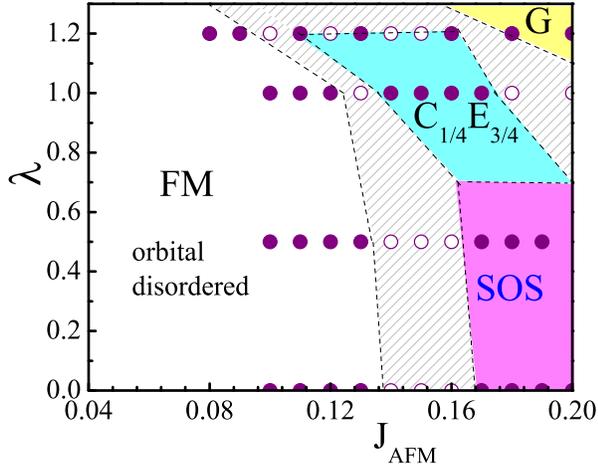}
\vskip -0.7cm
 \caption{(Color online) The zero-$T$ (2D lattice) phase diagram of quarter-doped manganites.
SOS denotes the novel multiferroic phase. Full dots denote the couplings where variational results were
confirmed by the MC simulation \cite{note} and the zero-$T$ relaxation using a $L=8$ cluster.
Open dots are cases where the MC simulation did not provide a clear answer due to strong
competition of metastable states.}
\vskip -0.65cm
\end{figure}

\textit{Results and Discussions}.
The 2D ground state phase diagram of the quarter-doped DE model is
shown in Fig.~1 \cite{3D}.
 A prominent FM metallic orbitally-disordered
phase occupies the small $J_{\rm AF}$ region,
in agreement with the phase diagram of large and intermediate
bandwidth  manganites. However, our most important result is revealed
 when $J_{\rm AF}$ is increased, keeping $\lambda$ small: in this regime
a new multiferroic phase (dubbed ``SOS'' as explained below)
is found. Its existence is confirmed by all the three
methods used here. In addition,
the C$_{1/4}$E$_{3/4}$ phase, predicted by Hotta \textit{et al.} \cite{Hotta:Prl},
appears when $\lambda$$\sim$1. Finally, a G-AFM region is found when
both $\lambda$ and $J_{\rm AF}$ are large enough \cite{comment-G}. Near the
boundaries  between these robust phases
the spin patterns are difficult to identify and phase separation
may exist there  (shaded regions in Fig.~1) \cite{comment-A}.
These regions are beyond our current computational capabilities.
Thus, here the emphasis will be only on the clearly identified states in the phase diagram.

\begin{figure}
\vskip -0.7cm
\includegraphics[width=0.50\textwidth]{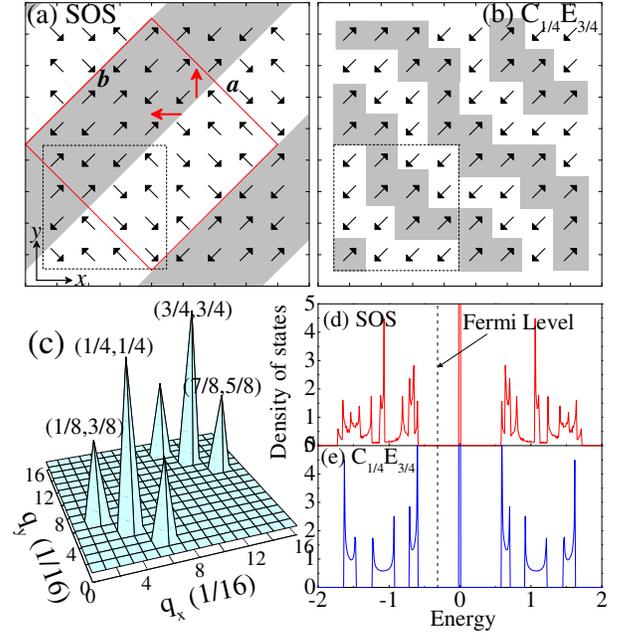}
\vskip -0.6cm
\caption{(Color online) Spin patterns of the (a) SOS and (b) C$_{1/4}$E$_{3/4}$ phases.
The magnetic unit cell of the SOS phase (red square) is $4\sqrt{2}\times4\sqrt{2}$
($4\times4$ in the orthorhombic notation). Shown also are the two possible
local $\vec{P}$ from
the DM mechanism (red arrows). The total $\vec{P}$ is, thus, perpendicular to the stripes~\cite{note3}.
(c) The (common) spin structure factor of these two phases.
The characteristic peaks appear at ($1/4$, $1/4$), ($1/8$, $3/8$), and ($3/8$, $1/8$) and their equivalent positions in the $\vec{q}$ + ($1/2$, $1/2$) region.
The ideal amplitudes of these three peaks are in the ratio of $2:1:1$. Density-of-states
of the (d) SOS and (e) C$_{1/4}$E$_{3/4}$  phases (at $\lambda=0$, fixing the spin pattern).
Both of these phases have a gap ($0.6$, $\sim0.2$-$0.3$ eV) at the Fermi level.}
\vskip -0.65cm
\end{figure}

Let us now analyze the physical properties of the SOS phase,
comparing properties  with those of the C$_{1/4}$E$_{3/4}$ phase.
The spin patterns of these two phases are shown in Fig.~2(a-b) (both
of them are perfectly AFM along the $c$-axis). For the SOS phase,
the most prominent features are the diagonal stripes where spin
domains with orthogonal orientations merge. For this reason, this
phase is called ``Spin-Orthogonal Stripe'' (SOS). This spin pattern
gives rise to a FE $P$ at the stripe boundaries via the
Dzyaloshinskii-Moriya (DM) interaction
($\vec{P}_{i,j}\propto\vec{e}_{i,j}\times(\vec{S}_{i}\times\vec{S}_{j})$,
where $\vec{e}_{i,j}$ is the unit vector connecting the NN spins
$\vec{S}_{i}$ and $\vec{S}_{j}$) \cite{Sergienko:Prb}. The two types
of local $\vec{P}$ that appear in the SOS state are shown in
Fig.~2(a). The total $\vec{P}$ arises from the sum of these local
$\vec{P}$, and it is perpendicular to the stripes. The local
$|\vec{P}|$ at the stripe boundaries should be larger than those in
TbMnO$_3$ and DyMnO$_3$, since here
$\textbf{S}_{i}\times\textbf{S}_{j}$ is the largest possible in
noncollinear spin orders. However, considering that only $25\%$ NN
spins are orthogonal, then the global $|\vec{P}|$ of the state
should be similar ($\sim10^3$ $\mu$C/m$^2$) as in TbMnO$_3$ and
DyMnO$_3$. In contrast, the C$_{1/4}$E$_{3/4}$ phase has all spins
collinear, forming FM zig-zag chains with alternative C- and E-AFM
sections~\cite{comment}.

Note that the spin structure factors ($S(\vec{q})=\sum_{i,\vec{r}}\vec{S}_{i}\cdot\vec{S}_{i+\vec{r}}\exp[i2\pi\vec{q}\cdot\vec{r}]$) of the SOS
and C$_{1/4}$E$_{3/4}$ phases are exactly the same (Fig.~2(c))~\cite{comment-same}.
The reason is that the spin pattern of the SOS phase can be viewed as the
combination of two C$_{1/4}$E$_{3/4}$ patterns:
by decomposing the spin vectors in Fig.~2(a) along the $x$ and $y$ axes, both
the $x$ and $y$ components by themselves form the C$_{1/4}$E$_{3/4}$ state,
with just a relative shift between the two C$_{1/4}$E$_{3/4}$ patterns formed by
this procedure.
Therefore,
care must be taken in separating the SOS and C$_{1/4}$E$_{3/4}$ states
experimentally because they are intimately connected.

It is also important to confirm that the SOS phase is an insulator,
as required by ferroelectricity. The density of states on
the infinite-size lattice indeed shows that both the SOS
and C$_{1/4}$E$_{3/4}$ phases have a gap at the Fermi level
(Fig.~2(d-e)). Thus, it is safe to predict that the SOS state
is a type-II multiferroic.

\begin{figure}
\vskip -0.6cm
\includegraphics[width=0.50\textwidth]{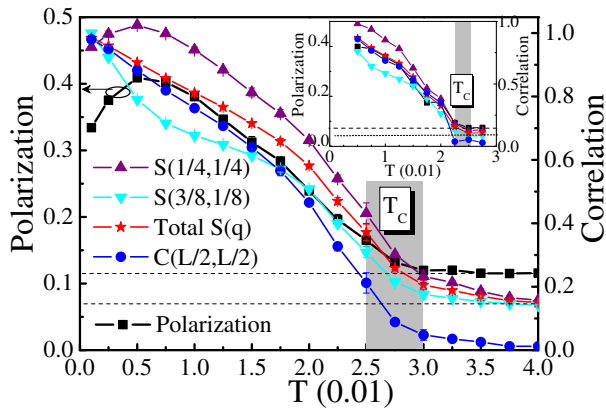}
\vskip -0.6cm \caption{(Color online) Finite-$T$ MC results ($J_{\rm
AFM}$=$0.18$, $\lambda$=$0$) for the SOS phase, on the $L$=$8$
cluster. {\it Left axis:} $\vec{P}$ per site is calculated using
$\sum_{<i,j>}[\vec{e}_{i,j}\times(\vec{S}_{i}\times\vec{S}_{j})]/L^2$,
phenomenologically. Thus, $|\vec{P}|$ per site at zero-$T$ is
$\sqrt{2}/4$. The MC average is done on $|\vec{P}|^2$, to avoid
strong finite-size fluctuations in $\vec{P}$, and
$\langle|\vec{P}|^2\rangle^{1/2}$ per site is shown. {\it Right
axis:} Amplitudes of characteristic peaks in the spin structure
factor and their summation are normalized to their zero-$T$ values.
Dashed lines indicate the size-dependent high-$T$ asymptotic values
of spin structure factors and $|\vec{P}|$ (i.e. background).
Real-space spin correlations at the maximum length $C$($L/2$, $L/2$)
are also shown. {\it Inset:} same properties but on $L$=$16$
lattices.} \vskip -0.6cm
\end{figure}

The theoretical prediction of the SOS phase is exciting
for its new mechanism for noncollinear spin ordering, which can improve the $T_{\rm C}$.
Traditionally, noncollinear spirals are formed by the competition between the NN and
next-nearest-neighbor (NNN) SEs, namely employing the SE
frustration \cite{Kimura:Prb,Dong:Prb08.2}. However, because the NNN SE
is usually weak, it becomes difficult to improve $T_{\rm C}$ based on such a
frustration mechanism. Another known path for noncollinear spin order
is the DM interaction \cite{Sergienko:Prb}, but it is also very weak and thus can only
induce small spin angles. In contrast, \emph{the $90^\circ$ spin angle in our SOS
phase is not driven by the NNN exchange frustration nor the DM interaction}.
Instead, the noncollinear spin structure of the SOS phase is driven
by the competition between DE and NN SE.
While in cuprate and nickelate stripes the spins across the stripes are AFM
ordered ($\pi$-shifted), in manganites there are also FM tendencies caused
by DE and, thus, a reasonable compromise at the stripes is to have $90^\circ$ spin angles.
Since both  DE and SE are NN interactions,
much stronger than NNN ones, a higher SOS $T_{\rm C}$ is expected.

To confirm this expectation, the finite-$T$ MC results, including the FE $|\vec{P}|$,
the spin structure factors, and the real-space spin
correlation at the maximum distance, are shown in Fig.~3. All these results
give a very similar $T_{\rm C}$, here defined as the $T$ where the magnetic correlation length
is as large as the clusters studied. For the $L$=$8$ lattice, the $T_{\rm C}$ of the SOS
phase is about $0.025$-$0.030$, which is $\sim$$3.85$-$4.6$ times the  $T_{\rm C}$
of the spiral state of previous investigations ($0.0065$,  obtained using the same
MC technique on a $L$=$12$ lattice \cite{Dong:Prb08.2}). Even for the $L$=$16$ lattice,
the SOS $T_{\rm C}$ is about $0.0225$-$0.025$, still much higher
than in the spiral state and showing that size effects are small. Therefore,
the real $T_{\rm C}$ of the SOS  phase is estimated
to be $\sim 100$ K ($4$ times the $T_{\rm C}$ of typical spiral manganites).
It is reasonable to expect the discovery of other
multiferroic noncollinear phases based on this electronic self-organization mechanism
with even higher $T_{\rm C}$'s.

\begin{figure}
\vskip -0.4cm
\includegraphics[width=0.45\textwidth]{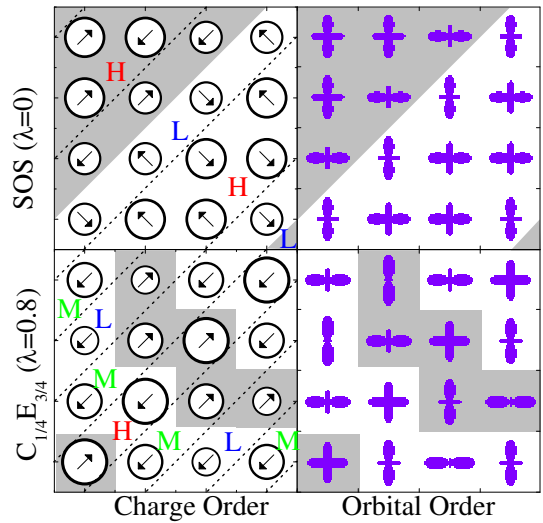}
\vskip -0.55cm
\caption{(Color online) Zero-$T$ charge/orbital ordering of the SOS and C$_{1/4}$E$_{3/4}$
phases. The unit cell of the charge/orbital ordering block shown here is only $1/4$ of
that of the  spin ordering (dotted regions in Fig.~2). The radius of the circles are
proportional to the local $e_{\rm g}$ charge density. The high (H), medium (M), and
low (L) charge density regions are partitioned by the broken lines. For the SOS phase
($\lambda=0$), the $e_{\rm g}$ density in the H (L) region is $5/6$ ($2/3$). For the
C$_{1/4}$E$_{3/4}$ phase at $\lambda=0.8$, the $e_{\rm g}$ density in the H/M/L sites
 is $0.95$/$0.73$/$0.59$, respectively.}
\vskip -0.7cm
\end{figure}

\textit{Charge/Orbital order and Coulomb interactions.}
Both the SOS and the C$_{1/4}$E$_{3/4}$ phases have a complex
charge/orbital order, forming stripes along the $b$-axis (see Fig.~4).
For the SOS phase, holes accumulate at the diagonal stripe
boundaries where $\vec{P}$ originates. For the C$_{1/4}$E$_{3/4}$ phase,
electrons accumulate at the E-AFM corner sites while
holes accumulate at the C-AFM bridge
sites~\cite{Hotta:Prl},
in agreement with the pure E-AFM (undoped case) and C-AFM
(over half-doped cases) properties. Also, note that the charge
disproportion in C$_{1/4}$E$_{3/4}$ is more prominent than that in the SOS phase.
Thus, if the Coulomb interaction is considered, the C$_{1/4}$E$_{3/4}$ phase
will be further
suppressed in the phase diagram. Using the mean-field approximation \cite{Hotta:Prb},
it can be shown that the phase boundary SOS-C$_{1/4}$E$_{3/4}$
can be shifted from $\lambda\approx0.7$ to $1$ by
incorporating the on-site Hubbard interaction. The orthorhombic distortion
(lattice shrinkage along the $c$-axis and
expansion in the $a$-$b$ plane)
will also  shift the phase boundary
SOS-C$_{1/4}$E$_{3/4}$ toward $\lambda\sim 1$.

The challenges for the experimental realization of the SOS phase are
substantial and will require much care: (1) The perovskite
manganites must be synthesized in the narrow-bandwidth limit. To
pursue such a large $J_{\rm AFM}/t_0$ ($>0.17$),
crystal with small lattice constants are needed (smaller than in
Pr$_{3/4}$Ca$_{1/4}$MnO$_3$ \cite{Kajimoto:Prb}).
(2) Another challenge is to avoid quenched disorder that may transform
long-range ordered phases
into glassy states \cite{Mufti:Prb,Subias:Prb}. These challenges may be solved by carefully
choosing divalent doping cations (not restricted to alkaline earths)
similar in size to those of small rare earths (e.g. Tb, Dy, and Ho). Also, recent
developments on complex oxide heterostrucutures may be helpful by creating
A-site ordered manganites \cite{Akahoshi:Prl} and controlling the crystal lattice
using strain/stress.

{\it Conclusions.} Here, a striped type-II multiferroic phase
of the two-orbital double-exchange model for quarter-doped manganites has been reported.
This phase is spin/charge/orbital ordered and has ferroelectricity due to the presence of
$90^\circ$ NN spin angles at spontaneously formed diagonal stripes.
This noncollinear spin-ordered state
potentially induces a multiferroic $T_{\rm C}$ higher than those of
spiral-spin ordered undoped manganites.

Work supported by the NSF (DMR-0706020) and the Division of
Materials Science and Eng., U.S. DOE, under contract with
UT-Battelle, LLC. J.M.L. was supported by the 973 Projects of China
(2009CB623303 and 2009CB929501) and NSF of China (50832002).


\begin{thebibliography}{bib}
\bibitem{note} Considering the orthorhombic distortion in real manganites, a weak NNN
SE ($J_{\rm 2b}$=0.01) along the $b$-axis was added in the MC simulation
to break the degeneracy between the $a$- and $b$-axes \cite{Dong:Prb08.2} to determine
the stripe direction. All identified phases' regions of stability are
virtually unaffected by $J_{\rm 2b}$. The single-ion anisotropy between
the $a$- and $b$-axes is not taken into account since it is even weaker
than $J_{\rm 2b}$  by $1$ or $2$ orders of magnitude. The MC simulations are initialized using randomly oriented spins, followed
by $1\times10^4$ MC steps (MCS) for thermalization, and $8\times10^3$ MCS for measurements.
$T=0.001$ for the MC simulation to approach ground state properties.
\bibitem{3D} Stacking the 2D states along the $c$-axis, and reversing the $t_{\rm 2g}$ spins
between adjacent planes (for the non-FM phases) due to the typical $c$-axis AFM coupling of insulating manganites,
the 3D variational phase diagram was found to be similar to Fig.~1.
\bibitem{comment-G} In our variational studies, the G-AFM state was assumed to have
the same staggered lattice distortions as the FM planes of the hole undoped A-AFM ground state.
\bibitem{note3} Since the spins' orientations are \emph{relative} in our models, due to
rotational invariance, the prediction of the direction of the DM
driven $\vec{P}$ for real manganites is not possible, with the
exception that it should be perpendicular to the stripe boundary.
Given the information about easy axes and planes for a specific
material, one can uniquely determine the direction of $\vec{P}$.
\bibitem{comment}
The mechanism proposed in \cite{Sergienko:Prl}
applied to the C$_{1/4}$E$_{3/4}$ phase can be shown to lead to a global $|\vec{P}|$=0.
\end{thebibliography}

\end{document}